\documentclass[final,1p,times]{elsarticle}

\usepackage{graphicx}
\usepackage{algorithmic}
\usepackage{algorithm}

\usepackage{amssymb}

\usepackage{subfigure}

\journal{Computer Communications (in press)}

\newcommand{\beq}{\begin{equation}}
\newcommand{\eeq}{\end{equation}}
\newcommand{\ba}[1]{\begin{array}{#1}}
\newcommand{\ea}{\end{array}}
\newcommand{\bea}{\begin{eqnarray}}
\newcommand{\eea}{\end{eqnarray}}

\newcommand{\ben}{\begin{enumerate}}
\newcommand{\een}{\end{enumerate}}
\newcommand{\bit}{\begin{itemize}}
\newcommand{\eit}{\end{itemize}}
\newcommand{\bde}{\begin{description}}
\newcommand{\ede}{\end{description}}

\newcommand{\ds}{\displaystyle}

\newcommand{\sz}{\scriptsize}

\begin{document}

\begin{frontmatter}



\title{Detecting communities of triangles in complex networks using spectral optimization}

\author[lyon]{Belkacem Serrour}
\ead{bserrour@bat710.univ-lyon1.fr}

\author[urv]{Alex Arenas\corref{cor}}
\ead{alexandre.arenas@urv.cat}

\author[urv]{Sergio G\'omez}
\ead{sergio.gomez@urv.cat}

\cortext[cor]{Corresponding author}

\address[lyon]{Lab.\ LIESP, B\^atiment Nautibus (ex 710), Universit\'e Claude Bernard Lyon 1, 69622 Villeurbanne Cedex, France}

\address[urv]{Departament d'Enginyeria Inform\`{a}tica i Matem\`{a}tiques, Universitat Rovira i Virgili, 43007 Tarragona, Spain}

\begin{abstract}
The study of the sub-structure of complex networks is of major importance to relate topology and functionality. Many efforts have been devoted to the analysis of the modular structure of networks using the quality function known as modularity. However, generally speaking, the relation between topological modules and functional groups is still unknown, and depends on the semantic of the links. Sometimes, we know in advance that many connections are transitive and, as a consequence, triangles have a specific meaning. Here we propose the study of the modular structure of networks considering triangles as the building blocks of modules. The method generalizes the standard modularity and uses spectral optimization to find its maximum. We compare the partitions obtained with those resulting from the optimization of the standard modularity in several real networks. The results show that the information reported by the analysis of modules of triangles complements the information of the classical modularity analysis.
\end{abstract}


\end{frontmatter}



\section{Introduction}

The study of the modular (or community) structure of complex networks has become a challenging subject \cite{firstnewman} with potential applications in many disciplines, ranging from sociology to computer science, see reviews \cite{jstat, schaeffer,lanfortunato}. Understanding the modular units of graphs of interactions (links) between nodes, representing people and their acquaintances, documents and their citation relations, computers and their physical or logical connections, etc., is of utmost importance to grasping knowledge about the functionality and performance of such systems. One of the most successful approaches to identify the underlying modular structure of complex networks, has been the introduction of the quality function called {\em modularity} \cite{newgirvan,newanaly}. Modularity encompasses two goals: (i) it implicitly defines modules as those subgraphs that optimize this quantity,  and (ii) it provides a quantitative measure to find them via optimization algorithms. It is based on the intuitive idea that random networks are not expected to exhibit modular structure (communities) beyond fluctuations \cite{rogerfluc}.

A lot of effort has been put into proposing reliable techniques to maximize modularity~\cite{newfast,clauset,donetti,santo,duch,amaral,latapy,pujol,newspect}, see review~\cite{fortunato1}. To a large extent, the success of modularity as a quality function to analyze the modular structure of complex networks relies on its intrinsic simplicity. The researcher interested in this analysis is endowed with a non-parametric function to be optimized: modularity. The result of the analysis will provide a partition of the network into communities such that the number of edges within each community is larger than the number of edges one would expect to find by random chance. As a consequence, each community is a subset of nodes more connected between them than with the rest of the nodes in the network. The user has to be aware of some aspects about resolution limitations that avoid grasping the modular structure of networks at low scales using modularity \cite{fortunato}. The problem can be solved using multiresolution methods \cite{bornholdt,nostre}.

The mathematical formulation of modularity was proposed for unweighed and undirected networks \cite{newgirvan} and generalized later to weighted \cite{newanaly} and directed networks \cite{sizered}. The generalized definition is as follows
\begin{equation}
Q\left(C\right) = \frac{1}{2w}\sum^{N}_{i=1}\sum^{N}_{j=1}\left(w_{ij} - \frac{w^{out}_{i}w^{in}_{j}}{2w}\right)\delta\left(C_{i},C_{j}\right)
\label{modularity}
\end{equation}
where $w_{ij}$ is the strength of the link between the nodes $i$ and $j$ of the network, $w^{out}_{i} = \sum_{j} w_{ij}$ is the strength of links going from $i$, $w^{in}_{j} = \sum_{i} w_{ij}$ is the strength of links coming to $j$, and the total strength of the network is $2w = \sum_{ij} w_{ij}$. Finally, $C_{i}$ is the index of the community to which node $i$ belongs to, and $\delta(x,y)$ is the Kronecker function assigning 1 only if $x=y$, and 0 otherwise.

A close look to Eq.(\ref{modularity}) reveals that the building block of the community structure we are looking for, within this formulation, is the link between two nodes. Every term in Eq.(\ref{modularity}) accounts for the difference, within a module, between the actual existence of a link with weight $w_{ij}$ and the probability of existence of such a link just by chance, preserving the strength distribution.

However, in many cases the minimal and functional structural entity of a graph is not a simple link but a small structure (motif) of several nodes \cite{motif-detec1}. Motifs are small subgraphs that can be found in a network and that correspond to a specific functional pattern of that network. Statistical over-representation of motifs (compared with the random occurrence of these sub-structures) has been a useful technique to determine minimum building blocks of functionality in complex networks, and several works exploit their identification \cite{motif-detec1,motif-detec3,motif-detec2}. Among the possible motifs, the simplest one is the triangle which represents the basic unit of transitivity and redundancy in a graph, see Figure~1. This motif is over-represented in many real networks, for example motifs~12 and~13 in Figure~1, the feedback with two mutual dyads and the fully connected triad respectively, are characteristic motifs of the WWW. Motif 7 (feed-forward loop) is over-represented in electronic circuits, neurons connectivity and gene regulatory transcription networks. The reason for this over-representation relies on the functionality of such small subgraphs on the evolution and performance of the specific network. In the WWW as well as in social networks, the fully connected triad is probably the result of the transitivity of contents or human relations, respectively. The feed-forward loop is related to the reliability or fail tolerance of the connections between important elements involved in communication chains. The idea we propose here is that finding modules containing such motifs as building blocks could improve our information about the modular structure of complex networks. The importance of transitivity is traced back to the seminal paper \cite{watts}
where it is proposed the clustering coefficient, a scalar measure quantifying the total number
of triangles in a network through the average likelihood that
two neighbors of a vertex are neighbors themselves.

The main goal of our work is to determine communities using as building blocks triangular motifs. We propose an approach for triangle community detection based on modularity optimization using the spectral algorithm decomposition and optimization. The resulting algorithm is able to identify efficiently the best partition in communities of triangles of any given network, optimizing their correspondent modularity function.

\begin{figure}[t]
	\centering
		\includegraphics[width=.90\textwidth]{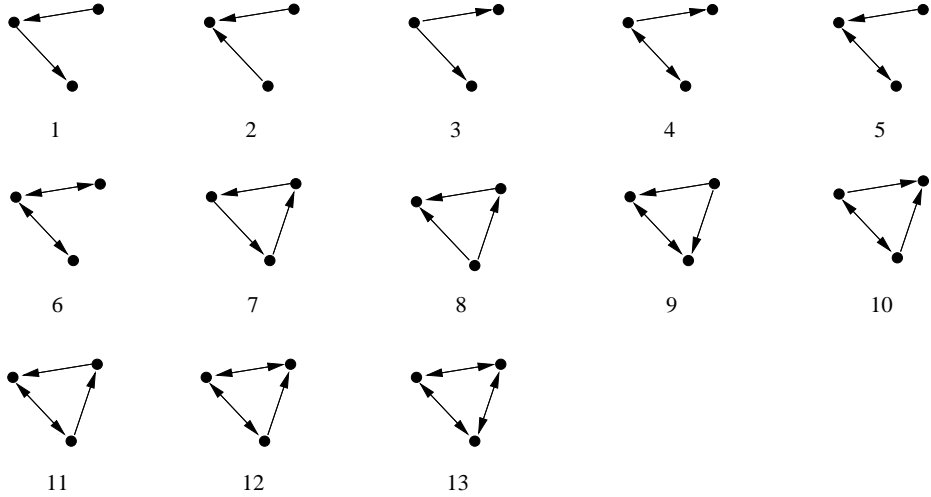}
		\caption{List of all possible three-nodes motifs.}
		\label{motif}
\end{figure}

\section{Spectral decomposition for triangle community detection}

Let $G=(V,A)$ be a weighted undirected graph representing a complex network, where $V$ represents the vertices set and $A$ the edges set. The objective is to identify communities of triangles, i.e.\ a partition with the requirement that the density of triangles formed by any three nodes $i$, $j$ and $k$ inside the same module is larger than the triangles formed outside the module. We will define this objective using a proper adaptation of modularity.

\subsection{Triangle modularity tensor}

In \cite{motifs} some of us introduced a mathematical formalism to cope with modularity of motifs of any size. Capitalizing on this work, here we study the specificity of triangle modularity $Q_{\triangle}(C)$ of a certain partition $C$ of an undirected graph (the extension to directed graphs is straightforward, although a little bit more intricate, we present this extension in the Appendix). The mathematical definition is
\beq
  Q_{\triangle}(C) =
    \sum_i \sum_j \sum_k B_{ijk} \delta(C_i,C_j) \delta(C_j,C_k) \delta(C_k,C_i)\,,
  \label{Qtriangles}
\eeq
where $C_i$ is the index of the community which node $i$ belongs to, and $B_{ijk}$
\beq
  B_{ijk} = \frac{1}{T_G}  w_{ij} w_{jk} w_{ki} -
            \frac{1}{T_N} (w_i w_j)
                          (w_j  w_k )
                          (w_k  w_i )
\eeq
is a three indices mathematical object ({\em triangle modularity tensor}, from now on) that evaluates for each triad $i$, $j$, $k$, the difference between the actual density of strength of the triangle in the graph and the expected density of this triangle in a random configuration with the same strength distribution (null case). The normalization constant $T_G$ is the total number of triads of nodes forming triangles in the network,
\beq
  T_G = \sum_i \sum_j \sum_k w_{ij} w_{jk} w_{ki}\,,
\eeq
and its counterpart $T_N$ for the null case term is
\beq
  T_N = \sum_i \sum_j \sum_k (w_i  w_j )
                             (w_j  w_k )
                             (w_k  w_i )\,.
\eeq
It is straightforward to check that the triangle modularity tensor satisfies:
\beq
  B_{ijk} = B_{jki} = B_{kij}\,,
  \label{bijkcycle}
\eeq
\beq
  \sum_i \sum_j \sum_k B_{ijk} = 0\,.
  \label{bijk0}
\eeq

\subsection{Spectral optimization of triangle modularity}

The computation of the triangle modularity is demanding due to the combinatorial number of triads that can be formed. The proposal of any optimization algorithm for this function must be aware of this cost. Among the possibilities already stated in the literature we devise that the spectral optimization scheme, first proposed in \cite{newspect}, is a candidate to perform this task efficiently. The idea behind this algorithm is to use the eigenspectrum of the modularity matrix, which plays a role in community detection similar to that played by the graph Laplacian, and use a recursion splitting reminiscent of graph partitioning calculations. The problem we have is that a direct mapping to the usual spectral modularity optimization is not straightforward given the structure of Eq.(\ref{Qtriangles}). Basically we need to transform Eq.(\ref{Qtriangles}) in a function with the following structure:
\beq
  Q(C) \propto \sum_i \sum_j s_i M_{ij} s_j\,,
\eeq
\noindent where the leading eigenvector of $M_{ij}$, the modularity matrix, will induce the first recursion step, splitting the network in two parts.

We propose the following transformation: let us assume a partition of the network in two communities, introducing the variables $s_i$, which are $+1$ or $-1$ depending on the community to which node $i$ belongs to, and taking into account that
\beq
  \delta(C_i,C_j) = \frac{1}{2}(1 + s_i s_j)\,,
\eeq
then
\bea
  \delta(C_i,C_j) \delta(C_j,C_k) \delta(C_k,C_i) & = &
    \frac{1}{8}(1 + s_i s_j) (1 + s_j s_k) (1 + s_k s_i) \nonumber \\
    & = & \frac{1}{4}(1 + s_i s_j + s_j s_k + s_k s_i)\,,
\label{deltas}
\eea
where we have made use of $s_i^2 = +1$. Therefore, using Eqs.~(\ref{bijkcycle}) and~(\ref{bijk0}),
\bea
  Q_{\triangle}(S) & = &
    \frac{1}{4} \sum_i \sum_j \sum_k B_{ijk} (1 + s_i s_j + s_j s_k + s_k s_i) \nonumber\\
    & = & \frac{3}{4} \sum_i \sum_j \sum_k B_{ijk} s_i s_j\,.
\eea
Defining the {\em triangle modularity matrix}
\bea
  M_{ij} & = & \sum_k B_{ijk} \nonumber\\
    & = & \frac{1}{T_G} w_{ij} \sum_k w_{jk} w_{ki} -
          \frac{1}{T_N} (w_i  w_i)
                          (w_j  w_j )
                          \sum_k (w_k  w_k )\,.
\eea
then
\beq
  Q_{\triangle}(S) = \frac{3}{4} \sum_i \sum_j s_i M_{ij} s_j\,.
\eeq
Thus, we have been able to reduce the optimization of the triangle modularity into the standard spectral algorithm given in~\cite{newspect}.

For the case of undirected networks, this matrix is symmetric and the computation of its eigenspectra gives real values. However, if the network is directed, this property is not necessarily true, and then a symmetrization of the matrix is needed before computing its spectrum (see Appendix).

Once a first division of the network in two parts has been obtained, it is possible to iterate the process, while modularity improves, by a recursive application of the spectral splitting to each subgraph. To this end, we need the value of the triangle modularity matrix for any subgraph. Supposing we have a subgraph $g$ to be divided into $g_1$ and $g_2$, the change in triangle modularity is given by
\bea
  \Delta Q_{\triangle}(g\rightarrow g_1,g_2) & = &
    \sum_{i,j,k\in g_1} B_{ijk} + \sum_{i,j,k\in g_2} B_{ijk} - \sum_{i,j,k\in g} B_{ijk} \nonumber \\
    & = &
    \frac{3}{4} \sum_{k\in g} \left(
      \sum_{i,j\in g} B_{ijk} s_i s_j - \sum_{i,j\in g} B_{ijk}
    \right) \nonumber \\
    & = &
    \frac{3}{4}\sum_{i,j\in g} s_i M_{ij}(g) s_j\,,
\eea
where
\beq
  M_{ij}(g) = \sum_{k\in g} \left( B_{ijk} - \delta_{ij}\sum_{\ell\in g} B_{i\ell k} \right)\,,
\eeq
and $s_i$ is $+1$ for nodes in $g_1$ and $-1$ for nodes in $g_2$. Therefore, the new triangle modularity matrix is not just a submatrix of the original one, but additional terms appear to take into account the connectivity with the rest of the network.

\subsection{Algorithm}

Once the triangle modularity has been transformed to the proper form to be optimized by spectral decomposition, we can proceed to formulate a complete decomposition-optimization algorithm. After the first analysis of the eigenspectra, the eigenvector associated to the largest eigenvalue is used to determine the elements that will be assigned to one of the two communties according to the sign of their eigenvector component. this process is recursively executed until no new splits are obtained. The decomposition given by the spectral partitioning can be improved by a fine-tuning of the nodes asignments after the process ends.

We use the Kernighan-Lin optimization method to improve the modularity as explained in \cite{newspect}. The main idea is to move vertices in a group to another increasing the modularity. We move all vertices exactly once. At each step, we choose to move the vertex giving the best improvement (largest increase in the modularity). When all vertices are moved, we repeat the process until no improvement is possible. Some computational issues should be considered here: the computation of the largest eigenvalue and its corresponding eigenvector can be efficiently determined using the iterative Lanczos method \cite{parlett}; the computation of $Q_{\triangle}(S)$ is, in principle, of order $O(N^3)$, however it can be done very efficiently by pre-computing and storing the values of $T_N$ and $T_G$, and the lists of triangles to which each node belongs to; finally, the KL post-processing stage which is eventually the computational bottleneck of the process, must be parameterized according to the number of nodes we pretend to move and the relative improvement of modularity observed.

\begin{algorithm}[t]
\algsetup{linenosize=\small}
 \begin{algorithmic}[1]
   \caption{Triangle community detection}
 \label{code}
   \REQUIRE \textit{Connected network G(V,E)}
   \ENSURE \textit{Triangle communities $C$, Triangle modularity of the partition $Q_{\triangle}(C)$}
   \STATE 	Read network
   \STATE 	Current subgraph $g$ $\leftarrow$ $G$
   \STATE 	Build modularity matrix $M(g)$
   \STATE   Compute $Q_{\triangle}(g)$
   \STATE   Compute leading eigenvalue and eigenvector of $M(g)$
   \STATE 	Decomposition of group $g$ in two groups: $g1$ and $g2$, using the signs of eigenvector components
   \STATE	  Compute the modularity $Q_{\triangle}(g1,g2)$ of the initial split of group $g$
   \STATE 	Improve $Q_{\triangle}(g1,g2)$ using KL optimization between $g1$ and $g2$
   \STATE  	Compute the modularity $Q_{\triangle}(g1,g2)$ of the split of group $g$
   \IF {$Q_{\triangle}(g1,g2)>Q_{\triangle}(g)$}
         \STATE {\bf goto} {\small 3} with $g$ $\leftarrow$ $g1$
         \STATE {\bf goto} {\small 3} with $g$ $\leftarrow$ $g2$
   \ENDIF
\end{algorithmic}
\end{algorithm}

\section{Results}

In this section we show the results of the algorithm, applied to several real networks. We have used the following networks:
\begin{itemize}
\item{Football \cite{firstnewman}, a network of American football games between Division IA colleges during regular season Fall 2000.}
\item{Zachary \cite{zachary}, a social network of friendships between 34 members of a karate club at a US university in the 1970s.}
\item{Dolphins \cite{lusseau}, an undirected social network of frequent associations between 62 dolphins in a community living off Doubtful Sound, New Zealand.} \item{Adjnoun \cite{newman06}, adjacency network of common adjectives and nouns in the novel David Copperfield by Charles Dickens.}
\item{Elec s208 \cite{motif-detec1}, benchmark of sequential logic electronic circuit.}
\item{Neurons \cite{worm}, network of neural connectivity of the nematode {\em C.elegans}.}
\item{Cortex \cite{scanell}, network of connections between cortical areas in the cat brain}.
\end{itemize}
To evaluate the information provided by the new triangle modularity, we perform a comparison with the standard modularity Eq.\ref{modularity}. We have developed a comparison in both the values of the optimal modularity, and the partitions obtained.

\subsection{Modularities comparison}

Table~\ref{trgComm} shows the best standard, and triangle modularities found using spectral optimization. We define a new parameter $\Delta(Q,Q_{\triangle})= (Q_{\triangle}-Q)/Q$ that measures the relative difference between both. Positive values of $\Delta(Q,Q_{\triangle})$ indicate that the contribution of triangles to communities is larger than standard modularity communities, and the contrary for negative values.

\begin{table}[tbp]
 \centering
   \begin{tabular}{lrrccc}
     \hline
     Network & Nodes & Links & $Q$ & $Q_{\triangle}$ & $\Delta(Q,Q_{\triangle})$ \\
     \hline
     Football & 115 & 613 & 0.604 & 0.924 & 0.529 \\
     Zachary & 34 & 78 & 0.419 & 0.706 & 0.685 \\
     Dolphins & 62 & 159 & 0.528 & 0.817 & 0.547 \\
     Adjnoun & 112 & 425 & 0.308 & 0.299 & -0.029 \\
     Elec s208 & 122 & 189 & 0.686 & 0.998 & 0.454 \\
     Neurons & 279 & 2287 & 0.405 & 0.433 & 0.069 \\
     Cortex & 55 & 564 & 0.372 & 0.708 & 0.903 \\
     \hline
   \end{tabular}
 \caption{Comparison of standard and triangle modularities.}
 \label{trgComm}
\normalsize
\end{table}

\begin{figure}[tpb]
 \begin{center}
 \begin{tabular}[t]{cc}
   \multicolumn{1}{l}{(a) Triangle modularity}
   &
   \multicolumn{1}{l}{(b) Standard modularity}
   \\ \\
   \mbox{\includegraphics*[width=.45\textwidth]{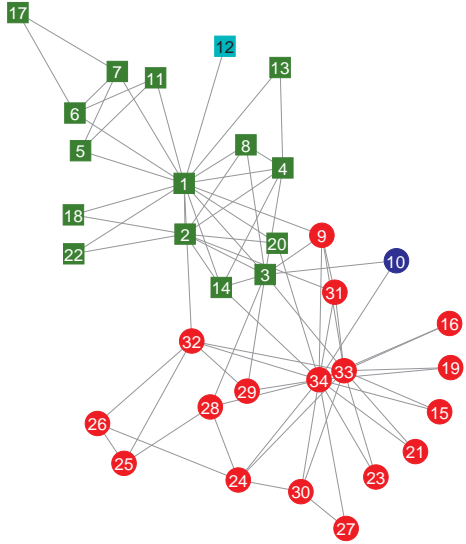}}
   &
   \mbox{\includegraphics*[width=.45\textwidth]{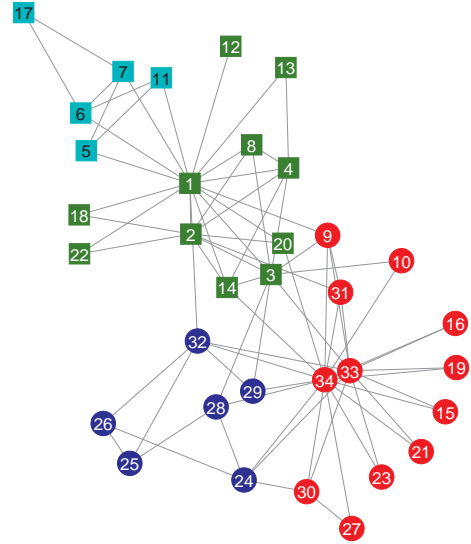}}
 \end{tabular}
 \end{center}
 \caption{Zachary network partitions. Best partitions found by optimization
of (a) triangle modularity and (b) standard modularity. The real splitting
of the network is represented by the shape of the symbols (squares and
circles). Colors indicate the assignment of nodes to the modules found.}
 \label{zacharyfig}
\end{figure}

From Table~\ref{trgComm} we observe that in Adjnoun, which is almost a bipartite network, the standard modularity is larger than the triangle modularity, in accordance with the absence of these motifs. On the other side, for the Zachary network, a human social network where transitivity is implicit in many acquaintances, the triangle modularity becomes more informative than the standard modularity. Indeed, the optimal standard modularity proposes a decomposition of this network in four groups, while the optimal triangle modularity  is achieved for a partition in two groups plus two isolated nodes (nodes 10 and 12) that do not participate in any triangle. Moreover the partition in two groups is in accordance with the observed split of this network after a fight between the administrator and the instructor of the club, see Figure \ref{zacharyfig}.

\subsection{Communities comparison}

A deeper comparison consist in to analyze the different modules obtained using the standard and triangle modularity. To this end, we need some measures to analyze the difference in the assignments of nodes to modules, taking into account that we will also have different modular partitions. Here, we use two measures, the Normalized Mutual Information (NMI) and the Asymmetric Wallace Index (AW).

In \cite{nmi} the authors define the NMI to compare two clusterings. The idea is the following: let be a clustering $A$ with $c_{A}$ communities and a clustering $B$ with $c_{B}$ communities, and let us define the confusion matrix $N$ whose rows correspond to the communities of the first clustering ($A$) and columns correspond to the communities of second clustering ($B$). The elements of the confusion matrix, $N_{\alpha\beta}$, represent the number of common nodes between community $\alpha$ of the clutering $A$ and community $\beta$ of the clustering $B$, the partial sums $N_{\alpha.}=\sum_{\beta}N_{\alpha\beta}$ and $N_{.\beta}=\sum_{\alpha}N_{\alpha\beta}$ are the sizes of these communities, and $N_{..}=\sum_{\alpha}\sum_{\beta}N_{\alpha\beta}$ is the total number of nodes. The measure NMI between two clusterings $A$ and $B$ is
\begin{equation}
\mbox{NMI}(A,B)=\frac{\ds
    -2\sum^{c_{A}}_{\alpha=1}\sum^{c_{B}}_{\beta=1}N_{\alpha\beta}
      \log\left(\frac{N_{\alpha\beta}N_{..}}{N_{\alpha.}N_{.\beta}}\right)
    }{\ds
    \sum^{c_{A}}_{\alpha=1}N_{\alpha.}\log\left(\frac{N_{\alpha.}}{N_{..}}\right)
    + \sum^{c_{B}}_{\beta=1}N_{.\beta}\log\left(\frac{N_{.\beta}}{N_{..}}\right)
    }
\label{eq3}
\end{equation}
If the partitions are identical, then NMI takes its maximum value of 1. If the partitions are totally independent, $\mbox{NMI} = 0$. It measures the amount of information that both partitions have in common.

The Asymmetric Wallace Index~\cite{wallace} is the probability that a pair of elements in one cluster of partition $A$ (resp.\ $B)$ is also in the same cluster of partition $B$ (resp.\ $A$). Using the same definitions as for the NMI, the two possible Asymmetric Wallace Indices are:
\begin{equation}
  \mbox{AW}_{1}(A,B) =
  \frac{\ds
    \sum^{c_{A}}_{\alpha=1}\sum^{c_{B}}_{\beta=1} N_{\alpha\beta} (N_{\alpha\beta} - 1)
  }{\ds
    \sum^{c_{A}}_{\alpha=1} N_{\alpha.} (N_{\alpha.} - 1)
  },
\end{equation}
\begin{equation}
  \mbox{AW}_{2}(A,B) =
  \frac{\ds
    \sum^{c_{A}}_{\alpha=1}\sum^{c_{B}}_{\beta=1} N_{\alpha\beta} (N_{\alpha\beta} - 1)
  }{\ds
    \sum^{c_{B}}_{\beta=1} N_{.\beta} (N_{.\beta} - 1)
  }.
\end{equation}
The asymmetric Wallace index shows the inclusion of a partition in the other.

\begin{table}[tbp]
 \centering
   \begin{tabular}{lcccc}
     \hline
     Networks &  NMI & AW$_1$ &  AW$_2$ \\
     \hline
     Football &  0.8903 & 0.8488 & 0.6901 \\
     Zachary &  0.6380 & 0.7945 & 0.5524 \\
     Dolphins &  0.6663 & 0.4810 & 0.7838 \\
     Adjnoun &  0.4888 & 0.3136 & 0.3845 \\
     Elec s208 &  0.6098 & 0.0307 & 0.9091 \\
     Neurons &  0.6045 & 0.7276 & 0.6954 \\
     Cortex &  0.8361 & 0.6841 & 1.0000 \\
     \hline
   \end{tabular}
 \caption{Comparison of partitions obtained using standard and triangles modularities. The different measures are explained in the text.}
 \label{metrics}
\normalsize
\end{table}

In Table~\ref{metrics}, we observe that the largest NMI is for the communities of football network. That means that the standard and triangle communities found in that network are very similar. Indeed, the structure of the football network is very dense and almost all nodes participate in triangles. For the the AW$_2$ of the cortex network is equal to 1, that means that all the triangle communities are included in the standard ones.

\section{Conclusions}

We have designed an algorithm to compute the communities of triangular motifs using an spectral decomposition of the triangle modularity matrix.
The algorithm provides partitions where transitive relations are the building blocks of their internal structure. The results of these partitions are complementary to those obtained maximizing the classical modularity, that accounts only for individual links, and can be used to improve our knowledge of the mesoscopic structure of complex networks.

\section{Appendix}

Here we show the computation of the triangle modularity matrix for a directed motif, in particular motif 7 in Figure~1, although as will be shown the process is equivalent for any other motif configuration. In this case, we have

\beq
  Q_{\triangle}(C) =
    \sum_i \sum_j \sum_k B_{ijk} \delta(C_i,C_j) \delta(C_j,C_k) \delta(C_k,C_i)\,,
\eeq
where $B_{ijk}$ is
\beq
  B_{ijk} = \frac{1}{T_G}  w_{ij} w_{jk} w_{ki} -
            \frac{1}{T_N} (w_i^{\mbox{\sz out}} w_j^{\mbox{\sz in}})
                          (w_j^{\mbox{\sz out}} w_k^{\mbox{\sz in}})
                          (w_k^{\mbox{\sz out}} w_i^{\mbox{\sz in}})\,.
\eeq
The normalization constant $T_G$ are now
\beq
  T_G = \sum_i \sum_j \sum_k w_{ij} w_{jk} w_{ki}\,,
\eeq
and
\beq
  T_N = \sum_i \sum_j \sum_k (w_i^{\mbox{\sz out}} w_j^{\mbox{\sz in}})
                             (w_j^{\mbox{\sz out}} w_k^{\mbox{\sz in}})
                             (w_k^{\mbox{\sz out}} w_i^{\mbox{\sz in}})\,.
\eeq

Using the transformation proposed in Eq.~(\ref{deltas})
\bea
  M_{ij} & = & \sum_k B_{ijk} \nonumber\\
    & = & \frac{1}{T_G} w_{ij} \sum_k w_{jk} w_{ki} -
          \frac{1}{T_N} (w_i^{\mbox{\sz out}} w_i^{\mbox{\sz in}})
                          (w_j^{\mbox{\sz out}} w_j^{\mbox{\sz in}})
                          \sum_k (w_k^{\mbox{\sz out}} w_k^{\mbox{\sz in}})\,.
\eea
then
\beq
  Q_{\triangle}(S) = \frac{3}{4} \sum_i \sum_j s_i M_{ij} s_j\,.
\eeq
Owing to the fact that the graph is directed, the modularity matrix $M_{ij}$ may be not symmetric, which causes technical problems. However, it is possible to restore the symmetry thanks to the scalar nature of $Q_{\triangle}(S)$~\cite{newmandirected}. A symmetrization of the triangle modularity matrix $M$,
\beq
  M^{'} = \frac{1}{2} (M + M^{T})\,,
\eeq
yields
\bea
  Q_{\triangle}(S) & = & \frac{1}{2} (Q_{\triangle}(S) + Q_{\triangle}(S)^{T}) \nonumber \\
   & = & \frac{3}{4} \sum_i \sum_j s_i M^{'}_{ij} s_j\,,
\eea
recovering the necessary symmetry to apply the standard spectral optimization.

In the same manner, we can define the modularity matrix for all possible motifs of Figure~\ref{motif} just by modifying $B_{ijk}$. For example, for motif~13 in Figure~\ref{motif} we have:

\bea
  B_{ijk} &=& \frac{1}{T_G}  w_{ij} w_{ji} w_{jk} w_{kj} w_{ki} w_{ik} \nonumber \\
  & &\mbox{}-
            \frac{1}{T_N} (w_i^{\mbox{\sz out}})^{2} (w_j^{\mbox{\sz in}})^{2}
                          (w_j^{\mbox{\sz out}})^{2} (w_k^{\mbox{\sz in}})^{2}
                          (w_k^{\mbox{\sz out}})^{2} (w_i^{\mbox{\sz in}})^{2}\,,\\
  T_G &=& \sum_i \sum_j \sum_k w_{ij} w_{ji} w_{jk} w_{kj} w_{ki} w_{ik}\,, \\
  T_N &=& \sum_i \sum_j \sum_k (w_i^{\mbox{\sz out}})^{2} (w_j^{\mbox{\sz in}})^{2}
                             (w_j^{\mbox{\sz out}})^{2} (w_k^{\mbox{\sz in}})^{2}
                             (w_k^{\mbox{\sz out}})^{2} (w_i^{\mbox{\sz in}})^{2}\,.
\eea

\section*{Acknowledgments}
We acknowledge J.~Borge-Holthoefer and A.~Fern\'andez for useful discussions. This work was supported by Spanish Ministry of Science and Technology FIS2009-13730-C02-02 and the Generalitat de Catalunya SGR-00838-2009. B.S. acknowledges support from he Rhone-Alpes region  for the financing of training by Explora'doc exchange scholarship.

\end{document}